\documentclass[twocolumn]{aastex63}
\usepackage{newtxtext,newtxmath}
\usepackage[T1]{fontenc}
\usepackage{ae,aecompl}
\usepackage{graphicx}	
\usepackage{amsmath}
	
\usepackage{amssymb}
\usepackage{float}
\usepackage[flushleft]{threeparttable}%to add a comment under tables
\usepackage{changepage}
\usepackage{epsf}
\usepackage{placeins}

\shorttitle{Catastrophic Outlier Photo-$z$ Identification}
\shortauthors{Singal et al.}
\reportnum{}

\begin{document}

\title{Machine Learning Classification to Identify Catastrophic Outlier Photometric Redshift Estimates}

\author{J. Singal}
\affiliation{Physics Department, University of Richmond, 
138 UR Drive, Richmond, VA 23173, USA\\}
\affiliation{Visiting Scholar, KIPAC, Stanford University, 382 Via Pueblo Mall, Stanford, CA 94305-4060, USA\\}

\author{G.~Silverman}
\affiliation{IT Architect, 
1429 9th Avenue \#1, San Francisco, CA 94122, USA\\}

\author{E. Jones}
\affiliation{Physics and Astronomy Department, University of California, Los Angeles, 
475 Portola Plaza, Los Angeles, CA 90095, USA\\}

\author{T. Do}
\affiliation{Physics and Astronomy Department, University of California, Los Angeles, 
475 Portola Plaza, Los Angeles, CA 90095, USA\\}

\author{B. Boscoe}
\affiliation{Physics and Astronomy Department, University of California, Los Angeles, 
475 Portola Plaza, Los Angeles, CA 90095, USA\\}
\affiliation{Computer Science Department, Occidental College, 
1600 Campus Road, Los Angeles, CA 90041, USA\\}

\author{Y. Wan}
\affiliation{Physics and Astronomy Department, University of California, Los Angeles, 
475 Portola Plaza, Los Angeles, CA 90095, USA\\}

\email{jsingal@richmond.edu}

\begin{abstract} 
We present results of using a basic binary classification neural network model to identify likely catastrophic outlier photometric redshift estimates of individual galaxies, based only on the galaxies' measured photometric band magnitude values.  We find that a simple implementation of this classification can identify a significant fraction of galaxies with catastrophic outlier photometric redshift estimates while falsely categorizing only a much smaller fraction of non-outliers.  These methods have the potential to reduce the errors introduced into science analyses by catastrophic outlier photometric redshift estimates.  
\\
\end{abstract}


\section{Introduction} \label{intro}

Upcoming large-scale surveys such as the Rubin Observatory Legacy Survey of Space and Time \citep[LSST -- e.g.][]{Ivezic08} which will observe up to hundreds of millions of individual galaxies in a limited number of photometric bands. Redshift determinations for most galaxies in these surveys will rely on photometric redshift estimation --- abbreviated ``photo-$z$'' --- see e.g. \citet{Salvato} for a recent review --- in which the redshifts are esimated from the fluxes in a small number of wavelength bins.  Many photo-$z$ estimation techniques are based on machine learning approaches, which work by developing a mapping from input parameters to redshift using a training set of galaxies whose redshifts are known, and then applying these mappings to an evaluation set for which the redshifts are to be determined.  

There are a fraction of galaxies that have a true redshift that is different from the estimated photo-$z$ redshift. These galaxies are often categorized into different degrees of ``outliers.'' One can define simple ``outliers''  as those galaxies where
\begin{eqnarray}
O: {{\vert z_{\rm phot}-z_{\rm spec} \vert} \over {1+z_{\rm spec}}} > 0.15,
\label{erroreq}
\end{eqnarray}
where $z_{\rm phot}$ and $z_{\rm spec}$ are the estimated photo-$z$ and actual (spectroscopically determined if available) redshift of the object, respectively  \citep[e.g.][]{Hildebrandt10,ML21}. For the most inaccurate photo-$z$ estimations the term ``catastrophic outliers'' (hereafter COs) is often used. Although there is not a standard, universal definition of COs, we use a definition that is typical \citep[e.g.][]{BH10,Graham18,WS21}:
\begin{eqnarray}
CO: {{\vert z_{\rm phot}-z_{\rm spec} \vert}} > 1.0.
\label{erroreqCO}
\end{eqnarray}
Those galaxies whose photo-$z$ estimates are sufficiently close to the actual redshift to not be outliers by the definition of equation \ref{erroreq} are termed ``non-outliers'' (NOs).  We note that because of the $1+z_{\rm spec}$ in the denominator of equation \ref{erroreq} it is technically possible for a galaxy with a redshift of \textasciitilde 6 or greater to be defined as a CO but also an NO. However this work does not include any galaxies for which this is possible.

\begin{table*}
\begin{center}
\scriptsize
\caption{Summary of test data set properties}
\label{tab0}
\begin{tabular}{ccccCC}
\hline
\hline
Name & Type & Number of Sources & $i$-band mag range (median) & \%Os & \%COs \\
\hline
`HSC' & spectroscopic redshifts + five-band photometry & 286,401 & 27.93-13.37 (19.63) & 10.9, 9.0 & 3.0, 5.1 \\
`COSMOS-reliable-$z$' & `quasi-spectroscopic' redshifts + five-band photometry & 58,619 & 27.11-19.00 (24.08) & 10.4, 9.0 & 1.9, 2.6 \\
\hline
\end{tabular}
\tablecomments{\footnotesize{Summary of test data sets used in this work, as discussed in \S \ref{tdata}.  Redshift distributions are shown in Figure \ref{z_dist}.   As discussed in the text, we use `quasi-spectroscopic' here to refer to an especially reliable photometric redshift derived from 30 filter bands extending from the infrared to the ultraviolet which we take as equivalent to a spectroscopic redshift.  The columns \%Os and \%COs are the percentage of outliers (as defined by equation \ref{erroreq}) and COs (as defined by equation \ref{erroreqCO}), respectively, for a validation set when photometric redshifts are determined with a) the basic neural network regression algorithm described in \S \ref{tdata} and b) the SPIDERz photo-$z$ code presented in \citet{Jones17}. }}
\end{center}
\vspace{0.2in}
\end{table*}

COs can have detrimental effects on the science goals of large scale surveys \citep[e.g.][]{BH10,Graham18}.  For example, in weak-lensing cosmic shear studies, estimations of cosmological parameters can begin to be affected with increased uncertainty and potential systematic errors if the percentage of COs is greater than 0.1 \citep{Hearin10}.  Therefore, mitigating COs is a crucial aim.  One promising strategy could involve a method of identifying or `flagging' potential catastrophic outliers, so that these galaxies could be excluded or deweighted in statistical analyses.  

In two previous works \citep{Jones18,WS21} we investigated the potential for identifying COs based on quantifiable features of their redshift probability distribution functions (PDFs) as output by a support vector machine learning method and found that such methods could correctly identify a significant fraction of COs while simultaneously incorrectly flagging a much smaller fraction of NOs. This method of CO identification is likely generalizable to other empirical redshift estimation methods but necessitates a redshift prediction method that develops a PDF in addition to a point photo-$z$ estimate.  Recently \citet{BG21} have analyzed using a neural network classifier to identify NOs with photometry plus select photo-$z$ confidence measures as input.

Here we investigate the question of whether a basic neural network binary classification algorithm can identify which galaxies are COs {\it based only on their photometric band values}, information that would be readily available in for any photometric redshift estimation method.  In this work we perform the initial photo-$z$ estimation with both a simple neural network regression algorithm and a support vector machine photo-$z$ estimator, although we emphasize that the results of this work are likely generalizable to photo-$z$s estimated with any empirical method.  

For the neural networks in this analysis we used architectures implemented in a standard way with Pytorch and run on Google Colab.  We present the network architecture specifications, and these methods are standard and can be implemented straightforwardly in Pytorch, TensorFlow, or any machine learning package.  Example Jupyter notebooks are available for the neural network binary classifier and the basic neural network regression photo-$z$ estimator.\footnote{https://sourceforge.net/projects/pzcobc/}

\begin{figure*} 
\begin{center}
\hspace*{-0.14in}
\includegraphics[width=2.3in,height=2.8in]{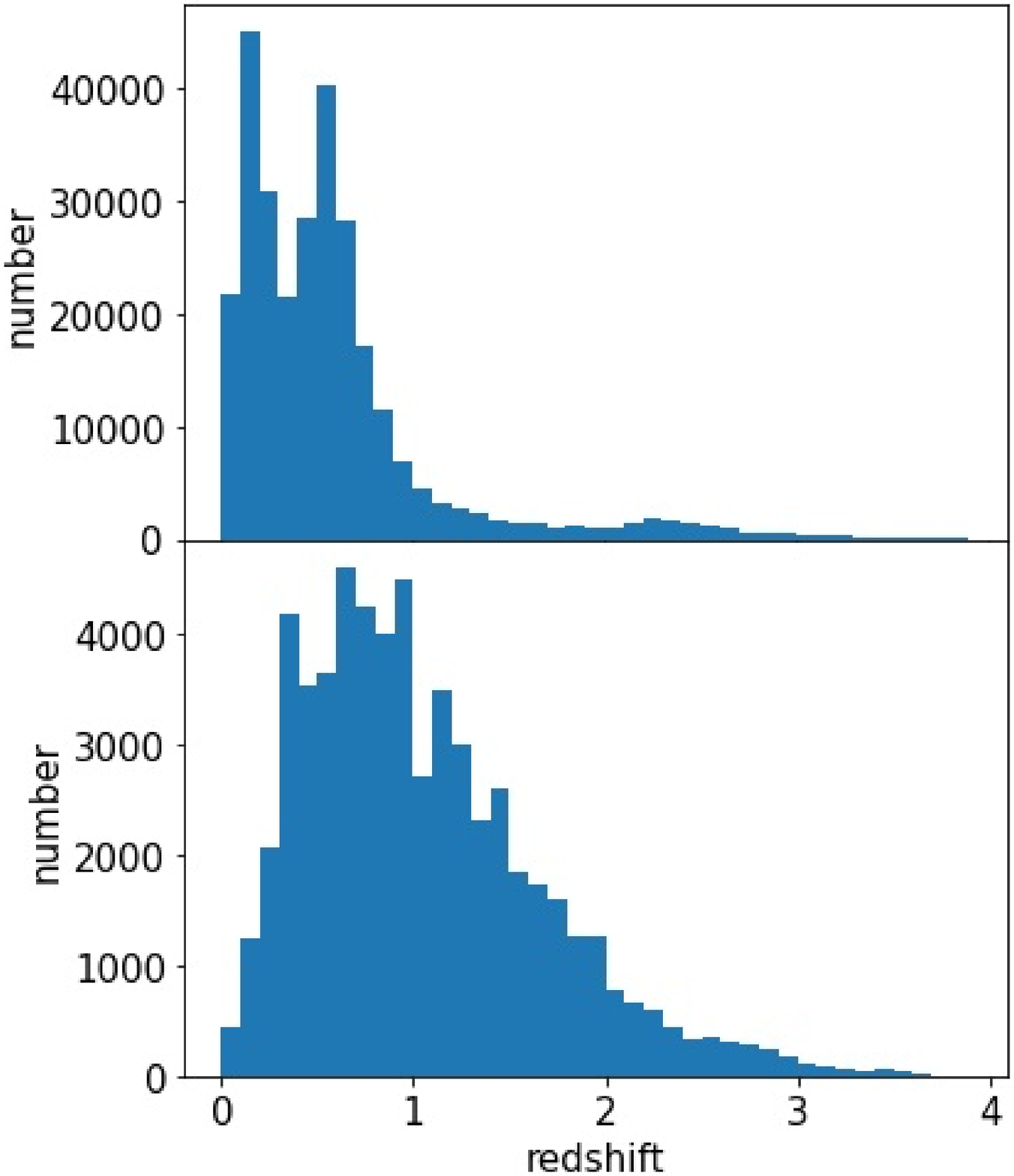}
\includegraphics[width=2.in,height=2.8in]{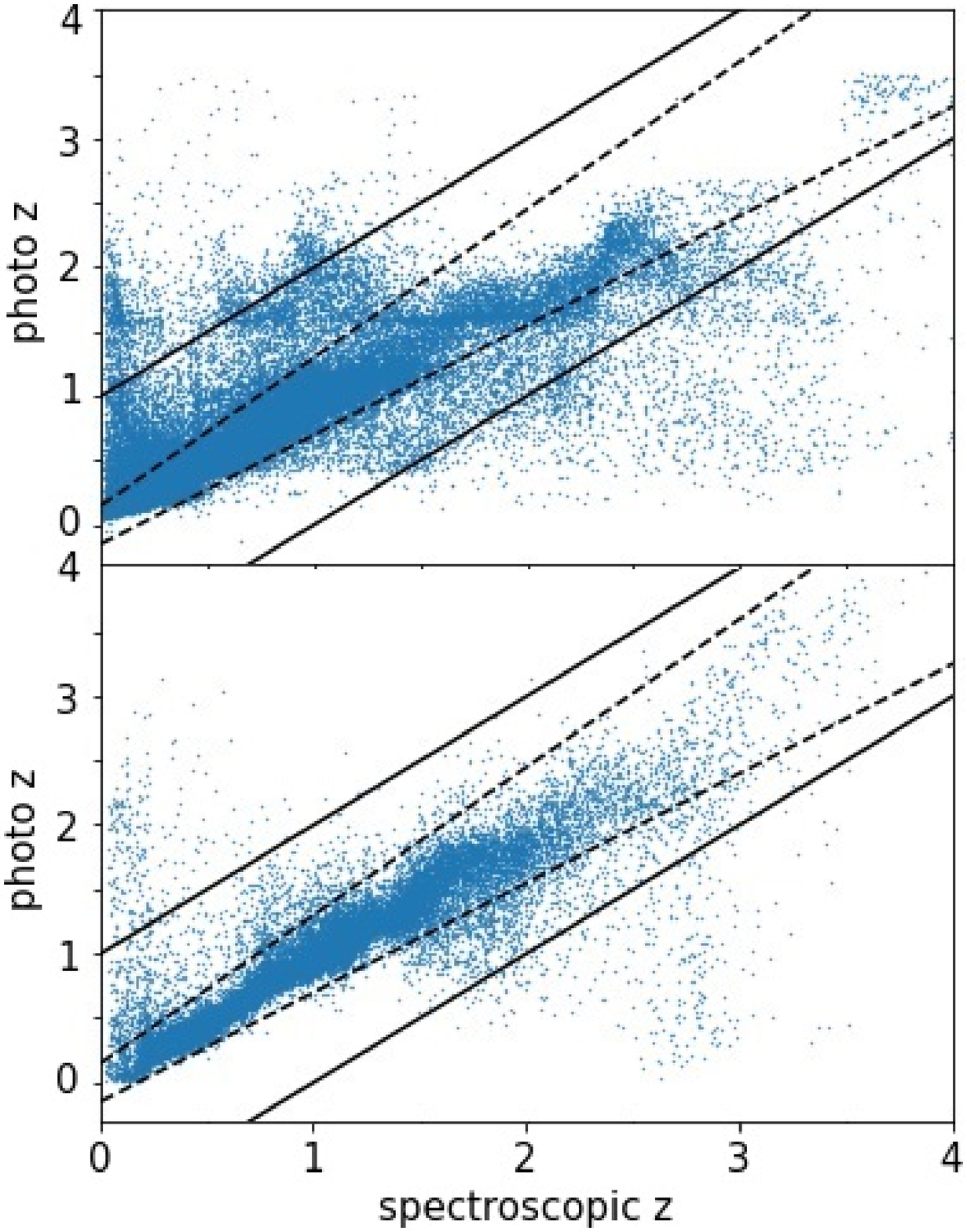}
\includegraphics[width=2.in,height=2.8in]{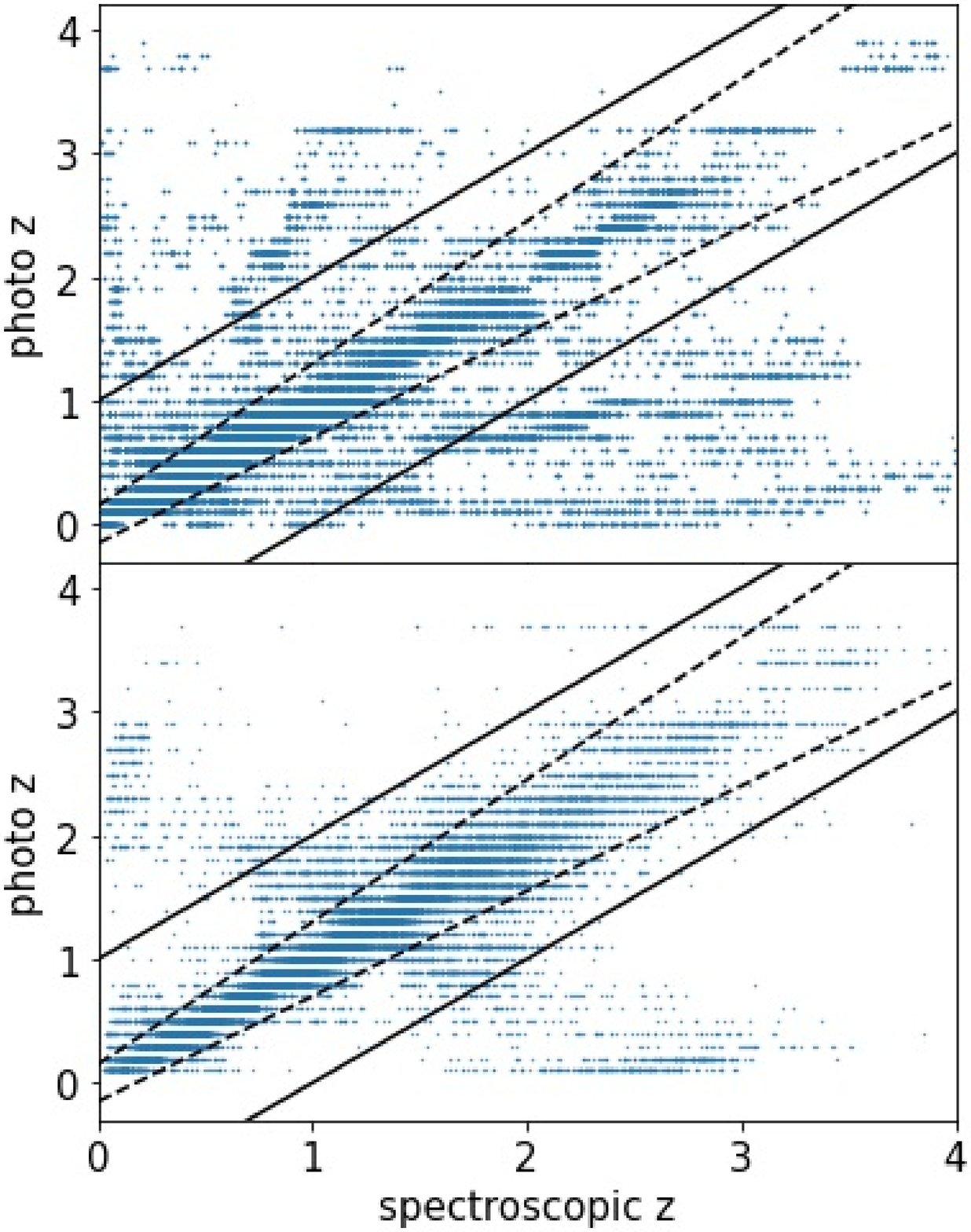}
\caption{The top row shows the HSC data set, and the bottom row shows the COSMOS-reliable-$z$ data set, which are the test data sets used in this analysis, both described in \S \ref{tdata}.  {\bf Left}: The redshift distribution $N(z)$ for the test data sets.  {\bf Center and Right}: Estimated vs. actual redshift for the validation sets of the test data sets as determined by a typical determination with the neural network regression model discussed in \S \ref{tdata} (center) and the SPIDERz photo-$z$ determination algorithm presented in \citet{Jones17} (right).  Those points outside of the inner diagonal (dashed) lines are outlier photo-$z$ estimates as defined by equation \ref{erroreq}, while those outside of the outer (solid) lines are CO photo-$z$ estimates as defined by equation \ref{erroreqCO}.  The vast majority of estimates are NOs as is typical of photo-$z$ estimations, and some statistics for these determinations are presented in Table \ref{tab0}.  Results for using the trained model to estimate redshifts of the training sets are similar.  These estimations are presented for completeness -- as discussed in \S \ref{tdata} the particular method or results of raw photo-$z$ estimation is not the major emphasis of this work.}
\label{z_dist}
\end{center}
\end{figure*}

In \S \ref{tdata} we present the test data sets used in this work.  In \S \ref{cato} we present the methods and quantification of results, and a discussion is given in \S \ref{disc}.

\section{Test Data} \label{tdata}

We require a test data set with a large number of galaxies with photometric data, as well as known redshifts over a wide range of redshifts and galaxy types.  We carry out this analysis with five optical photometric bands, both in order to approximate the conditions of a large scale survey such as LSST and because, as discussed below, a particularly large five band data set is available.  While five bands may seem overly pessimistic for LSST in particular, which will feature six photometric bands, many galaxies will not have overlap and will not have photometry in all six bands and/or will report large photometric errors in one or more bands \citep[e.g.][]{Soo18}.

One test data set we will use was compiled by \citet{Evan} and combines the Hyper-Suprime Cam (HSC) Public Data Release 2 \citep{aihara_second_2019}, which is designed to reach similar depths to LSST but over a smaller portion of the sky with cross-matched galaxy photometry from HSC with the HSC collection of publicly available spectroscopic redshifts from a number of works cataloged thoroughly in \citet{Evan}.  This data set contains photometry values for the bands $g$, $r$, $i$, $z$, and $y$ (we note that the photometric band $z$ is not to be confused with the redshift $z$ -- this is an unfortunate naming legacy).

For a second test data set with a different redshift distribution, we use the COSMOS2015 photometric catalog \citep{Laigle16} with a separate redshift estimation collection \citep{Ilbert09}, which contains particularly reliable redshifts derived from a very large number of photometric bands --- such a large number over such a large wavelength range that it approaches what we will term a `quasi-spectroscopic redshift.'  The COSMOS2015 catalog provides photometry for some galaxies in up to 31 optical, infrared, and UV bands, in addition to redshifts calculated from this photometry.  To maximize the reliability of the known quasi-spectroscopic redshifts for our photo-$z$ test purposes, we restrict the use of galaxies to those meeting the following criteria: (i) magnitude values present with no error flags for at least 30 bands of photometry, (ii) for which the ${\chi}^2$ for the \citet{Ilbert09} redshift estimate is $<1$, and (iii) for which the photo-$z$ value from the minimum ${\chi}^2$ estimate is less than 0.1 redshift away from the photo-$z$ value from the peak of the pdf. These galaxies can be considered to have highly reliable previous quasi-spectroscopic redshift estimates. Applying these criteria results in a data set of 58,622 galaxies. We then limited the redshifts to $z<4$ in order to prevent the occurrence of unoccupied bins, which resulted in a total of 58,619 galaxies.  We will refer to this set as `COSMOS-reliable-$z$.' For our test purposes  we restrict photo-$z$ estimation to the use of only five optical bands ($u$, $B$, $r$, $i$, and $z$). 

The redshift distributions of the test data sets are shown in Figure \ref{z_dist} and the parameters are summarized in Table \ref{tab0}.  We determine the photometric redshifts for these data sets in two ways.  The first is with a basic neural network regression model implemented in Pytorch, consisting of five hidden layers of 128 neurons each and one output neuron with rectified linear activation functions, with 200 training iterations of stochastic gradient descent back-propagation training optimization with a batch size of 32, for which we use 50\% of the objects for the training sets and 50\% for the validation sets.  Results of the raw point photo-$z$ estimations for this method are shown in Figure \ref{z_dist} for a typical determination.  For an alternate photo-$z$ determination method we use the SPIDERz support vector machine code presented in \citet{Jones17}.  As discussed in that work, we adopt the highest probability redshift for each galaxy as its point photo-$z$ estimate.  For the COSMOS-reliable-$z$ data set we use existing photo-$z$ estimations discussed in \citet{WS21}, while for the HSC data set we determine photo-$z$ values with new training and evaluation.  We emphasize that the particular methods of raw photo-$z$ estimation are not a major point of this work.  

Subsequent to determination of the basic photometric redshifts, we set aside a randomly chosen 30\% of the objects with predicted redshifts to serve as an evaluation set for the CO identification discussed below in \S \ref{cato}.  These objects are not used to train the CO binary classifier under any randomization.  We will refer to this set of objects as the ``base evaluation set.''  The base evaluation set thus consists of 85,952 objects for the HSC data set and 17,600 objects for the COSMOS-reliable-$z$ data set.

\section{Catastrophic Outlier Identification} \label{cato}

\begin{table*}
\begin{center}
\scriptsize
\caption{Metrics of performance for the CO binary classifier for a simple flagging with selected parameter values}
\label{tab2}
\begin{tabular}{cccccccc}
\hline
\hline
Test Data Set & Baseline Photo-$z$ Estimation & \% COs in Training & $w_{z>1}$ & $w_{z>2}$ & CO$_{\rm f}$\% & NO$_{\rm f}$\% & NO$_{\rm f}$\%>2.0 \\
\hline
HSC & NN regression & 35 & 7 & 150 & 37.2 & 2.8 & 21.6 \\
& SPIDERz & 25 & 2 & 30 & 35.1 & 3.4 & 6.2 \\
COSMOS-reliable-$z$ & NN regression & 50 & 55 & 350 & 34.9 & 3.9 & 38.1 \\
 & SPIDERz & 50 & 10 & 100 & 32.5 & 3.5 & 6.4\\
\hline
\end{tabular}
\tablecomments{\footnotesize{Some metrics of performance for the binary classifier neural network as discussed in \S \ref{cato} for a simple flagging where a value of greater than 0.50 in the output neuron results in the galaxy being flagged as a potential CO.  These are the results of the trained CO binary classifier evaluated on the original `base evaluation set' discussed in \S \ref{tdata}, choosing particular relatively well-performing instantiations of the training as discussed in \S \ref{cato}. The columns are as follows: `\%~COs in Training' is the percentage of COs in the modified ``rebalanced'' training set; $w_{z>1}$ and $w_{z>2}$ are the weight multiples applied to non-COs at redshifts $1\leq z<2$ and $z\geq 2$ respectively in the BCE loss function; CO$_{\rm f}$\% is the percentage of actual COs in the unaltered base evaluation set that are correctly flagged; NO$_{\rm f}$\% is the percentage of NOs that are incorrectly flagged; and NO$_{\rm f}$\%>2.0 is the percentage of NOs with an actual redshift greater than 2.0 which are incorrectly flagged; under this method.  The two test data sets and methods of baseline photo-$z$ determination are described in \S \ref{tdata}. As discussed in \S \ref{disc} these particular results do not necessarily represent an optimization over the parameters presented in the table or any others, but rather present a `proof of concept' of the use of such a binary classifier to flag potential COs.  }}
\end{center}
\end{table*}

Having photo-$z$ redshift estimates, we turn to the principal question of whether the photometry alone contains information that could identify a galaxy as a potential CO redshift estimate.  To address this question, we use the photometry along with the galaxies' known statuses as either a CO or not as input features to train a binary classification algorithm. For this binary classifier we implement a neural network with one output neuron with a sigmoid activation function and five hidden layers of 128 neurons each with rectified linear activation functions, in Pytorch, utilizing the Adam optimization of back propagation training with a batch size of 32 and Binary Cross Entropy (BCE) loss function.  

We find that training this network on the unmodified output of the photo-$z$ redshift estimations described in \S \ref{tdata} fails because the fraction of COs is too low (in the low single-digit percentages) so the network can achieve a minimal loss merely by classifying every object as a non-CO.  Instead, we modify the training sets to be skewed toward having larger fractions of COs, by simply removing at random a large number of non-COs.  This is similar to the training set modification known as ``deweighting'' which has been explored in the literature \citep[e.g.][]{AS,Hat}, and is similar in principle to the removal of lower redshift objects to achieve a higher proportion of high redshift objects that was explored in \citet{WS21} and which achieved more accurate point redshift estimates.  For clarity, due to other uses of the concept of weighting below, in this work we will refer to this particular training set modification with the alternate term of ``rebalancing.''  We find that this training set modification results in a good overall performance of identifying COs without misclassifying NOs but particularly misclassifies many NOs at redshifts $z>\sim1$. To remedy this, we apply weights to the BCE loss function (implemented in this case with the ``weight'' keyword in the Pytorch class BCEloss) during training such that objects that are at spectroscopic redshifts $1\leq z<2$ and $z\geq 2$ and that are non-COs are weighted a certain multiple times higher than other objects, thus more heavily penalizing misclassifying high redshift NOs.  We denote these weighting multiple parameters as $w_{z>1}$ and $w_{z>2}$ respectively.

We find that training for the binary classifier thus configured converges adequately with 200 iterations.  For this training we use 80\% training objects (from the rebalanced set).  A single training of the binary classifier with 200 iterations typically takes on the order of 10 minutes when running on a GPU on Colab.  We then apply the model trained on the rebalanced set to the `base evaluation set' that was set aside previously as discussed in \S \ref{tdata}.  This ensures that the performance of the binary classifier is evaluated on objects i) not used to train the binary classifier under any randomization and ii) with the (small) proportion of COs that actually result from the initial photo-$z$ determination.  There are several ways to make use of the output of the binary classifier.  The simplest and most straightforward use is to flag every galaxy with a value of greater than 0.50 in the  output neuron as a CO.  

We explored a variety of combinations of the weighting multiple parameters $w_{z>1}$ and $w_{z>2}$ and the percentage of COs in the rebalanced training set.  Table \ref{tdata} shows some metrics of performance for the trained binary classifier with this criterion, evaluated on the base evaluation sets with $\lesssim$5\% COs, for particular well-performing combinations of training set rebalancing modifications and loss function weighing multiple parameters.  We include the fraction of NOs that are incorrectly flagged both generally and for objects with $z>2$ as the latter can be viewed as more valuable given their relative rarity, as seen in the redshift determinations of Figure \ref{z_dist} (although it is also the case that for some science analyses objects with $z>2$ are not particularly valuable).  We find that the results when multiple realizations of trained models with the same parameters are applied to the base evaluation set are variable -- e.g. there is a range in the performance metrics reported in Table \ref{tdata}, and we choose particular instantiations of these models that have favorable results.  Such a freedom of choice would be available to any analysis that seeks to identify COs by this method in a real photo-$z$ analysis.  This is a point to which we will return in \S \ref{disc}.

\begin{figure*} 
\begin{center}
%\hspace*{-0.14in}
\includegraphics[width=1.70in,height=2in]{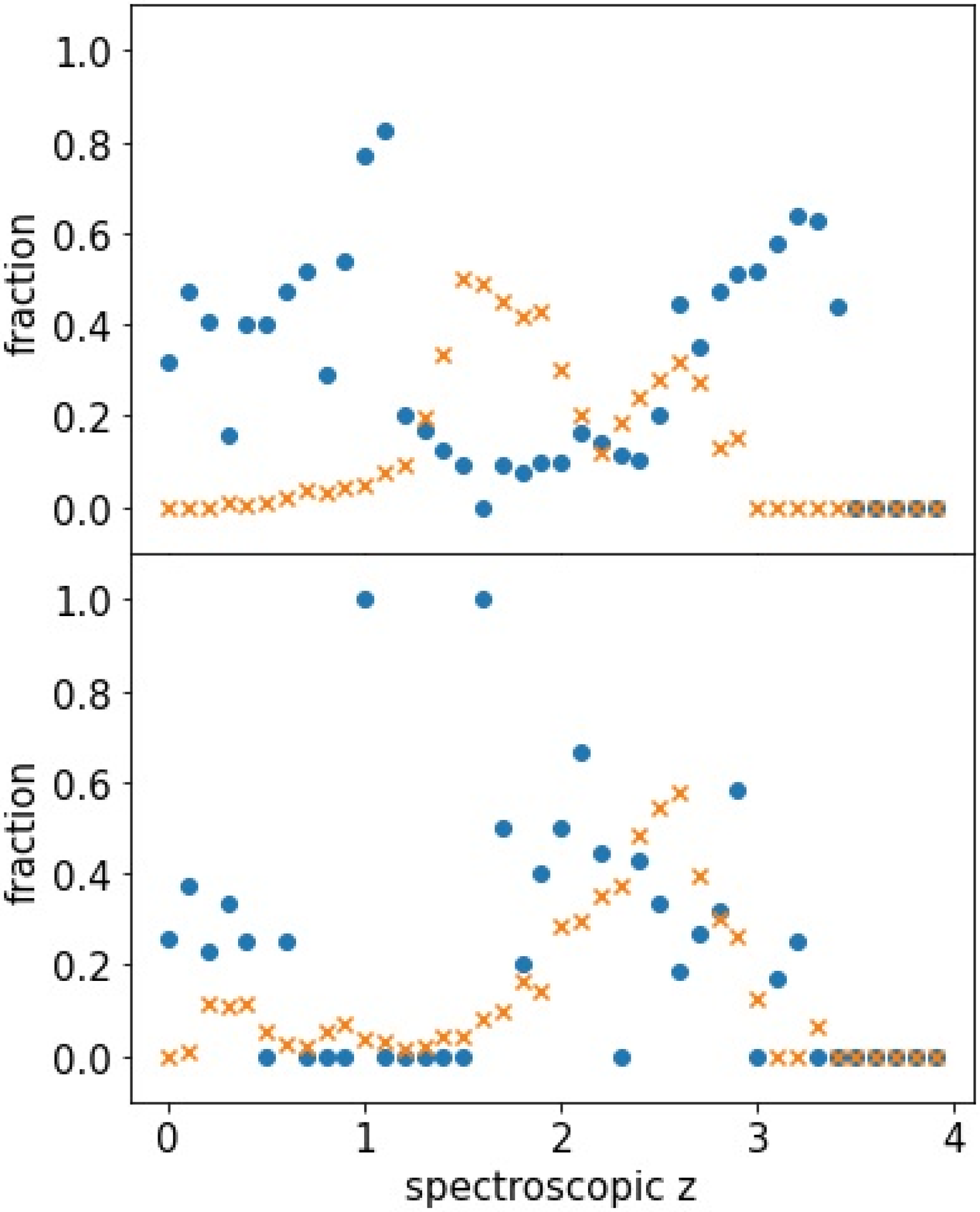}
\includegraphics[width=1.70in,height=2in]{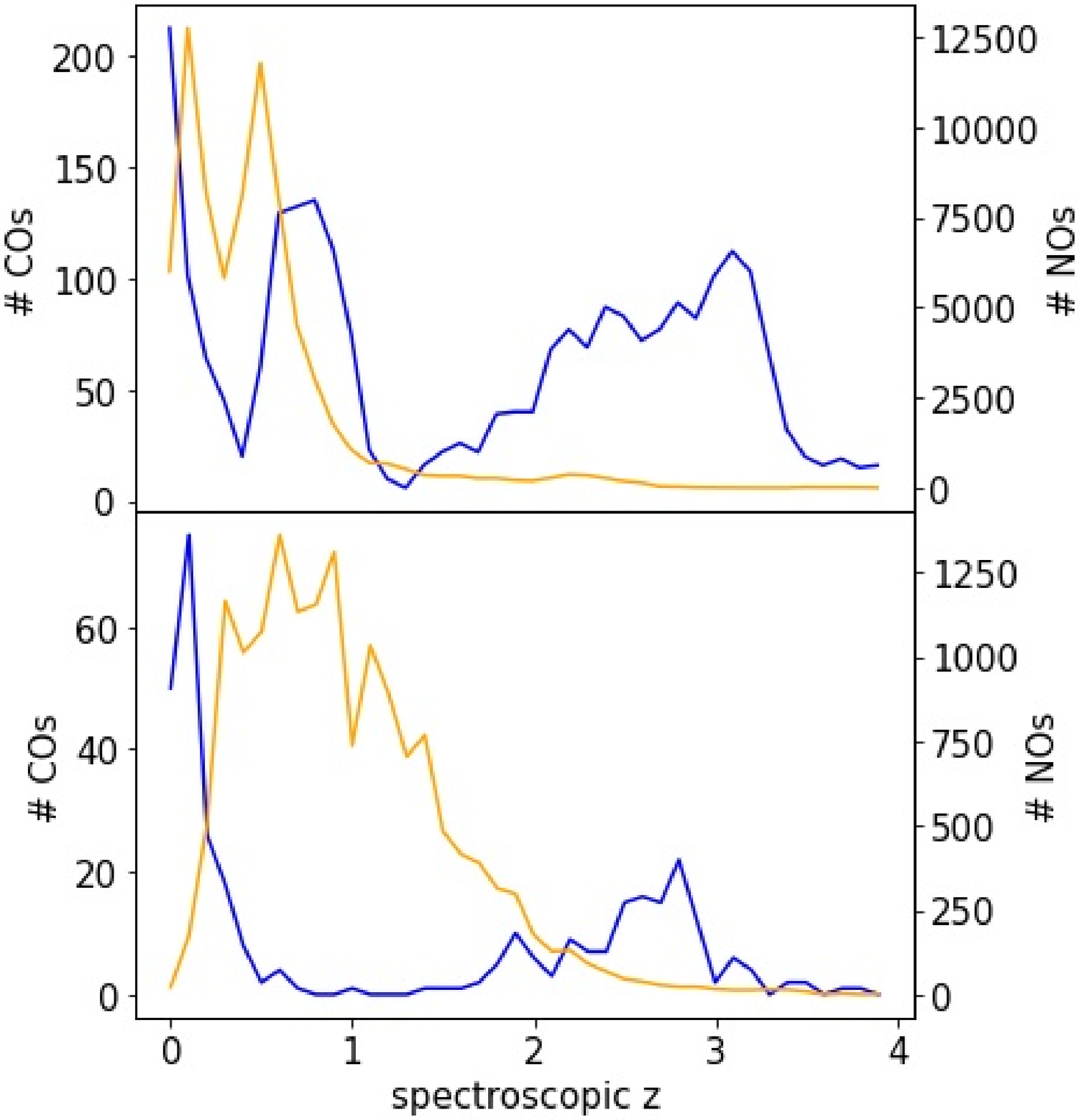}
---
\includegraphics[width=1.70in,height=2in]{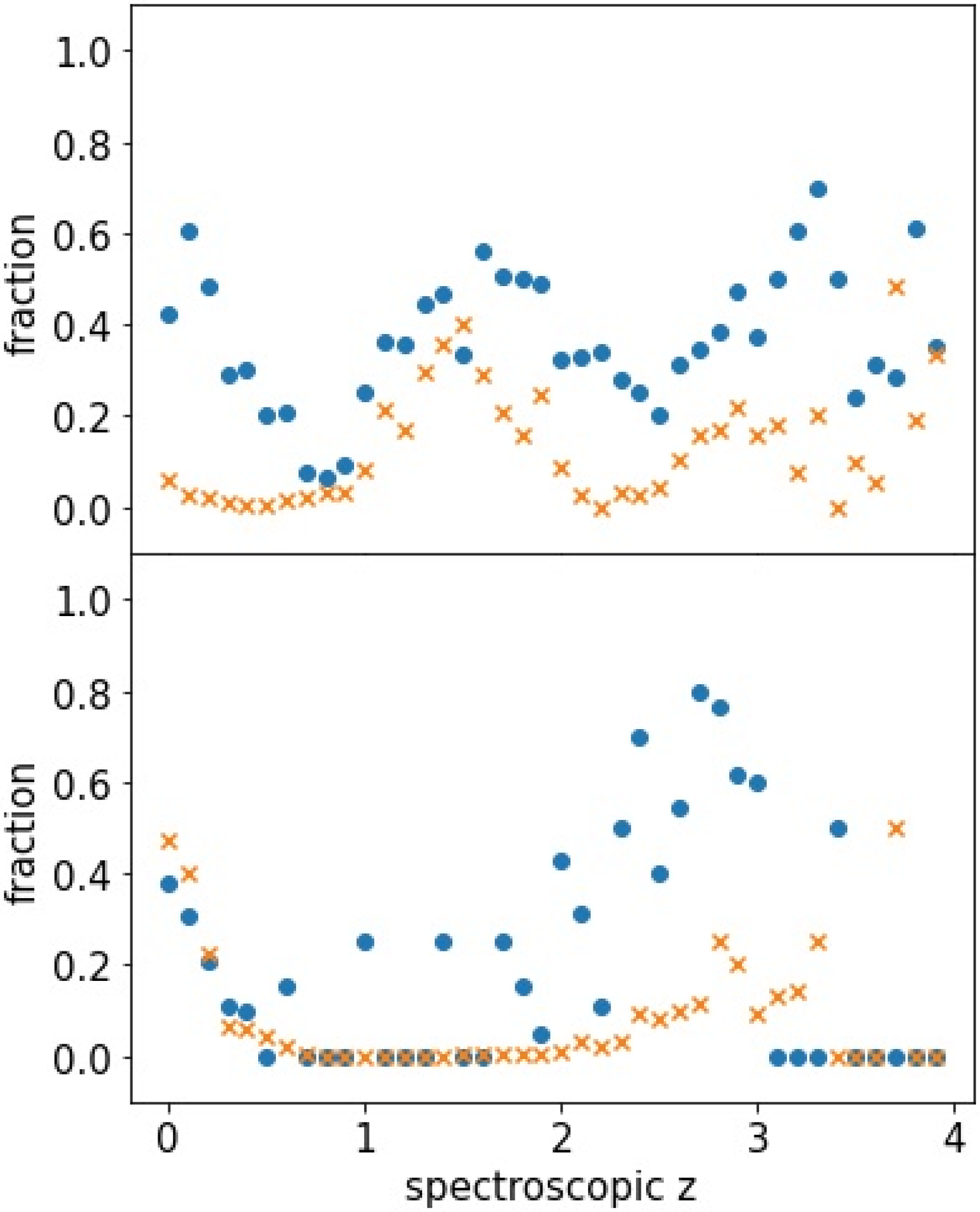}
\includegraphics[width=1.70in,height=2in]{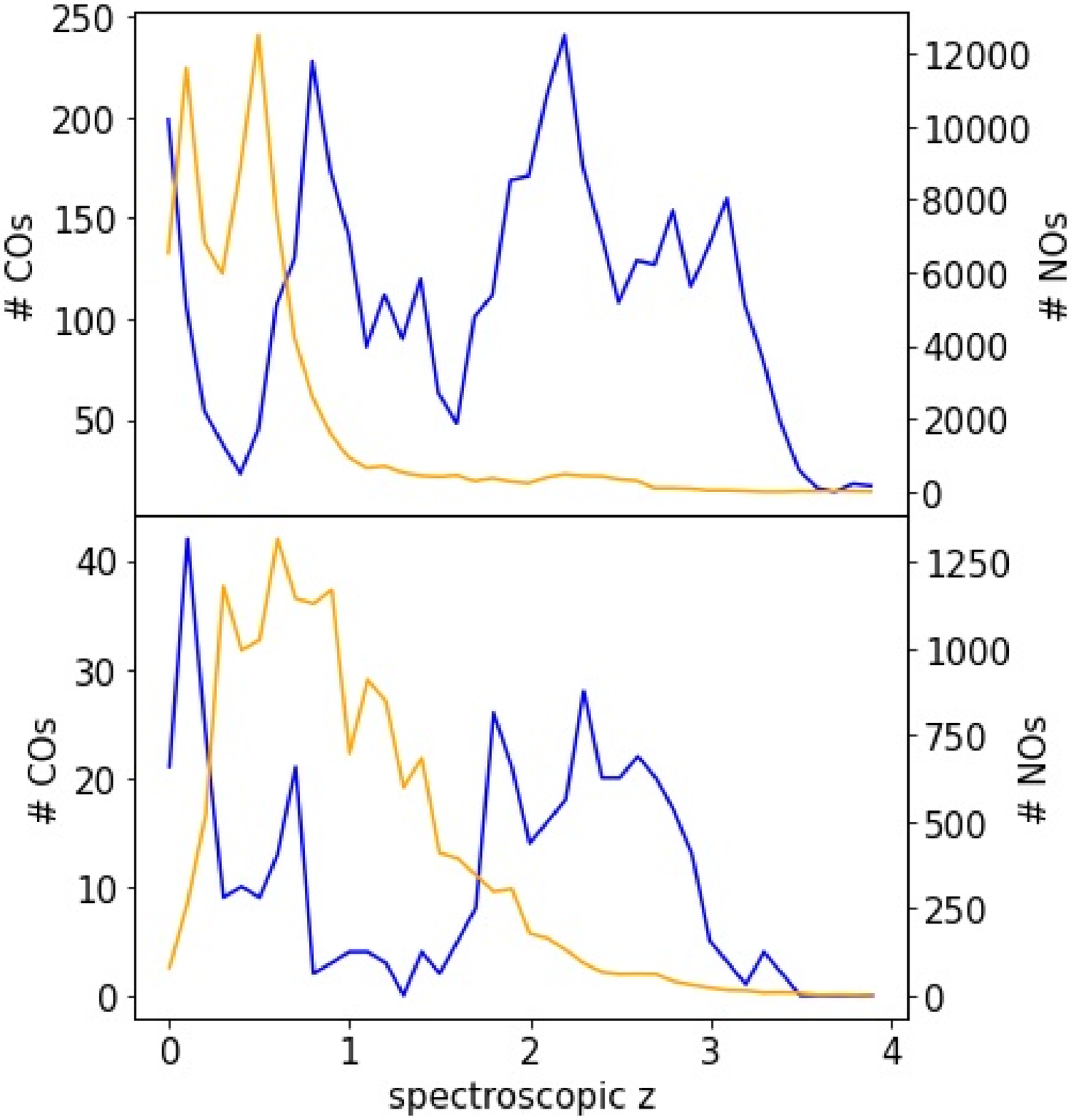}
\caption{The fraction of COs that are correctly flagged (blue circles) and NOs that are incorrectly flagged (orange xs) for the full unaltered base evaluation sets in bins of 0.1 in redshift, for a simple flagging where a value of greater than 0.50 in the output neuron results in the galaxy being flagged as a CO, as discussed in \S \ref{cato}.   Flagging results are shown for the HSC (top) and COSMOS-reliable-z (bottom) test data sets, for the relatively well-performing cases summarized in Table \ref{tab2}, where the total fractions (over all redshifts) of correctly flagged COs and incorrectly flagged NOs, as well as the parameters used, are presented.  Also shown are the total numbers in the base evaluation sets of COs (blue lines) and NOs (orange lines) in bins of redshift.  Results are shown for the Neural Network regression (left blocks) and SPIDERz support vector machine (right blocks) method of baseline photo-$z$ determinations.  Some redshift bins do not have many COs or NOs as seen in Figure \ref{z_dist}, which can lead to a very high or low fraction being flagged in those bins. As discussed in \S \ref{cato} and \S \ref{disc} different combinations of the tunable parameters of this analysis can lead to different results, and these results do not necessarily represent an optimization for any particular goal. }
\label{vsz}
\end{center}
\end{figure*}

Figure \ref{vsz} shows shows the fraction of COs that are correctly flagged and NOs that are incorrectly flagged as a function of redshift.  It can be seen that even when the overall fraction of incorrectly flagged NOs is small, the proportion in certain redshift ranges, usually at $z>1.5$ can be considerably higher, a characteristic that was also true of the CO identification strategy using redshift PDFs explored in \citet{WS21}.  

\section{Discussion} \label{disc} 

These results show that a simple binary classification neural network model can be implemented to correctly identify a significant fraction of CO photo-$z$ estimates while  incorrectly identifying a much lower fraction of NOs.  This has the potential to reduce the detrimental effects of CO photo-$z$ estimates in large-scale surveys if, for example, those galaxies flagged as COs were removed from, or weighted lower in, science analyses.  As discussed in \S \ref{cato} the efficacy of this binary classification relies on i) rebalancing the training set for the classifier to increase the fraction of COs and ii) implementing a weighting in the BCE loss function during training such that objects that are at spectroscopic redshifts $1\leq z<2$ and $z\geq 2$ and that are non-COs are weighted higher than other objects.

We emphasize that the purpose of this work is to demonstrate the efficacy of a simple implementation of a binary classification neural network to identify COs, and that we do {\it not} necessarily claim to have optimized the tunable parameters of this model architecture.  Such parameters include the fraction of COs in the rebalanced training set for the classifier, the factors $w_{z>1}$ and $w_{z>2}$ by which high redshift non-COs are more heavily weighted in the BCE loss function, the number of neurons and number of hidden layers, the loss function itself, the training batch size, the presence or absence of a skip layer, and others.  It is clear, however, given just the differences between the two test data sets and two baseline photo-$z$ determination methods considered here, that an optimization of any of these parameters would be entirely particular to any data set and any particular baseline photo-$z$ determination method, as well as the particular goal (e.g. maximally identifying COs vs. preventing loss of high redshift NOs).  For example, in weak lensing cosmic shear studies, the statistical importance of galaxies in the redshift ranges $z<0.3$ and $z>2.4$ is small \citep{Hearin10}, so in such studies losing proportionally more high redshift NOs in order to identify more COs in the redshift range $0.3\leq z \leq 2.4$ might be beneficial.  Likewise for cosmology with Type-Ia supernovae, LSST is unlikely to detect many Type-Ia supernovae in host galaxies with redshifts $z > 1.2$ \citep{ML21}, so for this science goal one may want to reduce COs even at the expense of eliminating many high redshift NOs.  If, on the other hand, the goal in question would be, for example, to identify specific high redshift sources for follow-up, then one would put a premium on preserving high redshift sources and therefore the avoidance of flagging of high redshift NOs. 

Any such optimization would have to be determined for any given photo-$z$ data set and any given photo-$z$ estimation method by setting aside from the original training set an equivalent of the base evaluation set used here and then testing different combinations of the tunable parameters.  Such a process could be automated with a grid search or carried out manually.  It is also important to note that because of the randomness inherent in selecting the 80\% training set for the binary classifier as well as the randomizing into batches of 32 for training, there is some variation in performance between iterations trained with identical values for these parameters.  However, once a trained network model is shown to perform well on a base evaluation set, one can be confident that it will perform well on the data set generally, as the base evaluation set is randomly chosen.  We note however that this is only absolutely true in the limit that the original training set is complete and representative of the galaxies for which photo-$z$ values are to be determined.  If that is not the case, then statistics of performance of a binary classifier on the base evaluation set are not guaranteed to be identical to those on the photo-$z$ evaluation set.  A possible mitigation for this would be to have a targeted spectroscopic campaign to achieve complete spectroscopic redshifts on a randomly selected sample of a few thousand objects from the overall photo-$z$ evaluation set and for that set to be used as the base evaluation set.

One may not necessarily even need to fully optimize over the set of parameter; rather, it may be useful to obtain a trained model that is `good enough' in terms of flagging a beneficial fraction of COs while not flagging a detrimental fraction of NOs, with the threshold for these specific to the survey or science goal in question.  We also especially note on this point that for some science analyses the relatively rarer galaxies at redshifts $z>2$ are valuable while for others they are not, and particularly in the case of the latter it may be quite beneficial to flag a large portion of COs that are actually at these redshifts but are falsely presenting as low redshift at the cost of flagging some high redshift NOs.

Even though any optimization of (or, as discussed above, obtaining a `good enough' result with) the tunable parameters of the binary classifier is particular to a given science goal, data set, and baseline photo-$z$ estimation method, it is the case that some general trends apply.  Increasing the percentage of COs in the training set and decreasing the threshold value for the output neuron both have the effect of increasing the proportion of correctly flagged COs along with increasing the proportion of incorrectly flagged NOs, while their inverses have the opposite effect.  For the data sets and photo-$z$ estimation methods considered here, both effects are very roughly linear (as opposed to, say, geometric) in the sense that, for example, changing the output neuron threshold value to 0.25 from 0.5 approximately doubles the fraction of correctly flagged COs and incorrectly flagged NOs.  The effects of the parameters $w_{z>1}$ and $w_{z>1}$ are more complicated.  Increasing their values lowers the portion of incorrectly flagged NOs at higher redshifts but also feeds back in more complicated ways, via the changes to the loss function and its minimization, to the proportion of correctly flagged COs and incorrectly flagged NOs at lower redshifts.  These effects are highly data set and photo-$z$ estimation method specific, as can be seen in the spread of these parameters' values which lead to favorable flagging results in e.g. Table \ref{tab2}.

It is the case that, as seen in Table \ref{tab2}, for both data sets and both baseline photo-$z$ estimation methods considered here, we were able to obtain results that were quite similar in the overall proportions of correctly flagged COs and incorrectly flagged NOs.  However, the results for the two photo-$z$ estimation methods differed substantially in the proportion of incorrectly flagged high redshift NOs, with that metric for the neural network regression being much higher than for SPIDERz, for both data sets.  This is very likely due to the different distributions of outliers that result from the two photo-$z$ estimation methods as can be seen in Figure \ref{vsz}; in particular, SPIDERz results in more and more extreme COs than the neural network regression.

The methods for identifying likely COs employed here enjoy one main advantage over the method using PDFs explored in \citet{Jones18} and \citet{WS21}: they can be implemented in the absence of PDF information and with any empirical photo-$z$ estimation method.   Several previous works \citep[e.g.][]{D08,MW08,ST13} also explored the use of probability information to potentially reduce the numbers of outliers and COs in photo-$z$ estimates of large survey data sets, a process sometimes referred to in those works as ``cleaning.''  A more thorough discussion of results from these works is presented in \citet{WS21} -- generally they achieve a lower efficiency of identifying COs and a greater misclassification of NOs than the present work or the PDF method of \citet{WS21}.  Recently \citet{BG21} have explored using a neural network classifier with photometry plus measures of photo-$z$ confidence to identify specifically NOs.  They find that a sample selected with this classifier that chooses the best third of the original set can achieve a significant reduction (30\%-40\%) in the fraction of outliers, although their analysis is restricted to redshifts $z \lesssim 1.6$. 

The results here indicate that at least a significant fraction of CO galaxies have distinct photometry patterns compared to NO galaxies in the same redshift range.  One could imagine several strategies to leverage this for identifying potential COs beyond the simple flagging based on a threshold value of the output neuron of the binary classifier incorporating only the photometric values exhibited here.  Such strategies could include incorporating, if available, redshift probability information, or galaxy shape information from imaging, into the inputs to the binary classifier, or using a generative adversarial network or some other architecture.  It is our intention to explore some of these in future works.  The present work serves as a proof of concept that useful CO identification with a classification algorithm is possible.

\section*{Acknowledgements}

HSC data are provided by the NAOJ / HSC Collaboration.  This work is based in part on observations made with the NASA/ESA Hubble Space Telescope, obtained from the Data Archive at the Space Telescope Science Institute, which is operated by the Association of Universities for Research in Astronomy, Inc., under NASA contract NAS 5-26555.

\end{document}